\documentclass[fleqn,11pt,twoside]{article}

\usepackage{amsthm,amsthm,amssymb, color, xcolor,epsfig, graphics, subfigure}

\usepackage{amsmath, graphicx, latexsym, lscape}

\usepackage{epic,epsfig}
\usepackage{graphicx}
\usepackage[all]{xy}
\usepackage{tikz}
\usepackage{mathtools}
\usepackage{mdframed}

\usepackage{placeins}

\usepackage{ifthen}

\newcommand{\xw}[3]{%
  \ifthenelse{\equal{#3}{777777}}{{\node at (#1,#2) {{$.$}};}}{
  \ifthenelse{\equal{#3}{888888}}{{\node at (#1,#2) {{$\circ$}};}}{
  \ifthenelse{\equal{#3}{999999}}{{\node at (#1,#2) {$\bullet$};}}
  {\node at (#1,#2) {$#3$};}}}}

\newcommand{\xlew}[3]{%
  \ifthenelse{\equal{#3}{777777}}{{\node at (#1*1.38,#2) {{$.$}};}}{
  \ifthenelse{\equal{#3}{888888}}{{\node at (#1*1.38,#2) {{$\circ$}};}}{
  \ifthenelse{\equal{#3}{999999}}{{\node at (#1*1.38,#2) {$\bullet$};}}
  {\node at (#1*1.38,#2) {$#3$};}}}}

\newcommand{\stenhmi}[2]{{
   \draw (#1,#2) circle (0.25cm);
    \draw (#1+2,#2+2) circle (0.3cm);
    \draw (#1+1,#2+1) circle (0.25cm);
    \draw (#1+1,#2) circle (0.25cm);
   \draw (#1+1,#2+2) circle (0.25cm);
    \draw (#1+2,#2+1) circle (0.25cm);
    \draw (#1,#2+1) circle (0.25cm);
}}

\newcommand{\stenbsq}[2]{{
   \draw (#1,#2) circle (0.3cm);
    \draw (#1+2,#2) circle (0.25cm);
    \draw (#1+2,#2+2) circle (0.25cm);
    \draw (#1+1,#2+1) circle (0.25cm);
    \draw (#1+1,#2) circle (0.25cm);
    \draw (#1+1,#2+2) circle (0.25cm);
    \draw (#1+2,#2+1) circle (0.25cm);
    \draw (#1,#2+1) circle (0.25cm);
    \draw (#1,#2+2) circle (0.25cm);
}}

\newcommand{\stenhmn}[3]{{
    \draw (#1*#3,#2) circle (0.2cm);
    \draw (#1*#3+2*#3,#2+2) circle (0.2cm);
    \draw (#1*#3+1*#3,#2+1) circle (0.2cm);
    \draw (#1*#3+1*#3,#2) circle (0.2cm);
    \draw (#1*#3+1*#3,#2+2) circle (0.2cm);
    \draw (#1*#3+2*#3,#2) circle (0.2cm);
    \draw (#1*#3,#2+2) circle (0.2cm);
}}

\newcommand{\bse}{\begin{subequations}}
\newcommand{\ese}{\end{subequations}}

\newcommand{\be}{\begin{equation}}
\newcommand{\ee}{\end{equation}}
\newcommand{\bea}{\begin{eqnarray}}
\newcommand{\eea}{\end{eqnarray}}
\newcommand{\nn}{\nonumber}

\newcommand{\beq}{\begin{equation}}
\newcommand{\eeq}{\end{equation}}

\def \CP2{{$\mathbb CP^2$}}

\makeatletter
\newcommand{\copyrightnote}[2]{{\renewcommand{\thefootnote}{}
 \footnotetext{\small\it
\begin{flushleft}
 \copyright \ #1   #2  
\end{flushleft}}}}

\newcommand{\Name}[1]{\begin{flushleft}
                       \LARGE \bf #1
                       \end{flushleft}\vspace{-3mm}}

\newcommand{\Author}[1]{\begin{flushleft}
                       \it #1 \end{flushleft}}

\newcommand{\Address}[1]{\begin{flushleft}
                       \it #1 \end{flushleft}}

\newcommand{\Date}[1]{\begin{flushleft}
                      \small  \it #1 \end{flushleft}}

\newcommand{\evenhead}{Author \ name}
\newcommand{\oddhead}{Article \ name}

\renewcommand{\@evenhead}{
\hspace*{-3pt}\raisebox{-15pt}[\headheight][0pt]{\vbox{\hbox to \textwidth
{\thepage \hfil \evenhead}\vskip4pt \hrule}}}
\renewcommand{\@oddhead}{
\hspace*{-3pt}\raisebox{-15pt}[\headheight][0pt]{\vbox{\hbox to \textwidth
{\oddhead \hfil \thepage}\vskip4pt\hrule}}}
\renewcommand{\@evenfoot}{}
\renewcommand{\@oddfoot}{}

\long\def\@makecaption#1#2{%
  \vskip\abovecaptionskip
  \sbox\@tempboxa{\small \textbf{#1.}\ \ #2}%
  \ifdim \wd\@tempboxa >\hsize
    {\small \textbf{#1.}\ \ #2}\par
  \else
    \global \@minipagefalse
    \hb@xt@\hsize{\hfil\box\@tempboxa\hfil}%
  \fi
  \vskip\belowcaptionskip}

\newcommand{\JNMPnumberwithin}[3][\arabic]{%
  \@ifundefined{c@#2}{\@nocounterr{#2}}{%
    \@ifundefined{c@#3}{\@nocnterr{#3}}{%
      \@addtoreset{#2}{#3}%
      \@xp\xdef\csname the#2\endcsname{%
        \@xp\@nx\csname the#3\endcsname .\@nx#1{#2}}}}%
}

\newcommand{\resetfootnoterule} {
  \renewcommand\footnoterule{%
  \kern-3\p@
  \hrule\@width.4\columnwidth
  \kern2.6\p@}
}

\renewcommand{\footnoterule}{}

\makeatother

\theoremstyle{definition}

\begin{document}

\renewcommand{\evenhead}{ {\LARGE\textcolor{blue!10!black!40!green}{{\sf \ \ \ ]ocnmp[}}}\strut\hfill J Hietarinta}
\renewcommand{\oddhead}{ {\LARGE\textcolor{blue!10!black!40!green}{{\sf ]ocnmp[}}}\ \ \ \ \  Degree growth of lattice equations
defined on a 3×3 stencil}

\thispagestyle{empty}
\newcommand{\FistPageHead}[3]{
\begin{flushleft}
\raisebox{8mm}[0pt][0pt]
{\footnotesize \sf
\parbox{150mm}{{Open Communications in Nonlinear Mathematical Physics}\ \ \ \ {\LARGE\textcolor{blue!10!black!40!green}{]ocnmp[}}
\quad Special Issue 1, 2024\ \  pp
#2\hfill {\sc #3}}}\vspace{-13mm}
\end{flushleft}}

\FistPageHead{1}{\pageref{firstpage}--\pageref{lastpage}}{ \ \ }

\strut\hfill

\strut\hfill

\copyrightnote{The author(s). Distributed under a Creative Commons Attribution 4.0 International License}

\begin{center}
{  {\bf This article is part of an OCNMP Special Issue\\ 
\smallskip
in Memory of Professor Decio Levi}}
\end{center}

\smallskip

\Name{Degree growth of lattice equations\\ defined on
  a $3\times3$ stencil}

\Author{Jarmo Hietarinta}

\Address{Department of Physics and Astronomy,\\
    University of Turku, FIN-20014 Turku, Finland}

\Date{Received July 14, 2023; Accepted January 11, 2024}

\setcounter{equation}{0}

\begin{abstract}

\noindent 
We study complexity in terms of degree growth of one-component lattice
equations defined on a $3\times 3$ stencil.  The equations include two
in Hirota bilinear form and the Boussinesq equations of regular,
modified and Schwarzian type. Initial values are given on a staircase
or on a corner configuration and depend linearly or rationally on a
special variable, for example $f_{n,m}=\alpha_{n,m}z+\beta_{n,m}$, in
which case we count the degree in $z$ of the iterates. Known
integrable cases have linear growth if only one initial values
contains $z$, and quadratic growth if all initial values contain
$z$. Even a small deformation of an integrable equation changes the
degree growth from polynomial to exponential, because the deformation
will change factorization properties and thereby prevent
cancellations.
\end{abstract}

\label{firstpage}


{\it Dedicated to the memory of Decio Levi.}

\section{Introduction}
The concept of integrability is associated with the dynamics being
regular (as opposed to chaotic), without being simple. Since the
equations are nonlinear, regularity means there must be some
underlying ``controlling'' mathematical structures.  For example,
solutions to integrable equations are often associated with elliptic
functions.  Since many different types of (nonlinear) equations can
show regular behavior there cannot be a strict all-encompassing
definition of integrability.

It is more fruitful to consider ``indicators'' of integrability, each
with their own range of applicability. For example, the three-soliton
condition (3SC) is a good indicator of integrability for partial
differential or partial difference equations in Hirota bilinear form, while
the Painlev\'e property is applicable more widely for differential
equations but not so easily for difference equations. These two are
algorithmic ways to prove integrability, that is, following a given
procedure one can prove or disprove integrability; usually it is
easier to disprove.  Certain other integrability indicators
require the construction of additional structures, such as conserved
quantities, or Lax pairs, or B\"acklund transformations etc. Such
constructions may require a lot of ingenuity.

In this paper, we only consider integrability in the context of partial
difference equations (for an overview, see \cite{HJN,GHRV}). For them
one algorithmic and powerful method is the study of the growth of
complexity under iterations, characterized by ``algebraic entropy''
\cite{Ve92,FaVi93,BeVi99,HV,HV2,HV3,HaPr05,Via15}.  We will briefly discuss
this in Section \ref{C:ent}.

Growth properties of 2D one-component equations defined on a single
quadrilateral of the $\mathbb Z^2$ lattice have been studied by
several authors, see e.g. \cite{TrGrRa01,Viallet2006,Mase15,
  RobertsTran19,HMW,UWGR}.  If the starting configuration is on the
quadrilateral $(0,0) - (1,1)$ of the Cartesian lattice and the
evolution is to the NE direction, then one can look for factorization
and cancellation first at the point $(2,2)$ (and this was used as a
condition in the search for possible integrable equations in
\cite{HV2d}). As for lattice equations defined on a larger than
$2\times2$ stencil, some results exist for the Toda lattice, defined
on a star shaped 5-point stencil \cite{KMT15}.

The main results of this paper concern the numerical analysis of degree
growth for partial difference equations defined on a
$3\times 3$ stencil. The methodology is discussed in Section
\ref{C:comp}.  In Section \ref{C:H}, we will first study two equations
in Hirota bilinear form, one integrable and one non-integrable, which
depend on seven point of the $3\times 3$ stencil. These equations are
relatively simple and we can study integrability from various points
of view, in order to establish the validity of the computational
method.  Then in Section \ref{C:B} ,we analyze one-component equations
of Boussinesq type, which involve all nine points in the $3\times 3$
stencil.

\section{Degree growth and cancellations\label{C:ent}}
When one studies discrete dynamics (maps), one way to quantify the
idea of ``regularity'' is to study the complexity of their
iterates. This association was made already in the 1990's by Veselov
\cite{Ve92} and others.  When a rational map is iterated, its
complexity can be quantified as the degree of the computed numerator
(or denominator) with respect to the initial values. In general, the
degrees grow exponentially with the number of iterations $n$ but the
growth will be reduced if the numerator and denominator have a common
factor which can be canceled. This was analyzed in detail in
\cite{FaVi93,BeVi99,HV,HV2,HV3,HaPr05} and it was found that for
integrable maps the cancellations are strong enough to convert the
exponential degree growth to polynomial growth with respect to $n$. In
fact the conjecture relating degree growth (after cancellations) to
integrability is as follows:
\begin{itemize}
\item If the degree growth is linear in $n$ then the equation is
  linearizable.
\item If the degree growth is polynomial in $n$ then the equation is
  integrable.
\item If the degree growth is exponential in $n$ then the equation is chaotic.
\end{itemize}

In order to observe the cancellations, it is best to formulate a higher
order map as a multi-component first order map in projective
space. Then, instead of a rational expression, we have multi-component
polynomial maps, and cancellations take place when, after some number
of iterations, the components have a nontrivial common factor.

The existence of a common factor after $n$ iterations means also that
the initial values, for which the common factor vanishes, are singular
points, because starting from those points the iterations eventually
lead outside the projective space.  Thus cancellations and
singularities are  two sides of the same phenomenon.

Associated to this is also the method of ``singularity confinement''
\cite{GRP-PRL,GRWM,Mase_2019}.  In this method one studies the
singularity further by starting from a point infinitesimally close to
the singular point and checking whether after passing the singularity
the dynamics becomes regular eventually. This is a very effective
method and been used extensively in order to de-autonomize discrete
maps.

Still another approach is to study the singularity itself using methods from
algebraic geometry, namely ``resolution of singularities''
\cite{Takenawa2001a,Takenawa2001c}.

\section{Lattice equations and their initial data}
Among the integrable partial differential equations we have first
order evolution equations like the Korteweg -- de Vries (KdV) equation
and also second order equations like the Boussinesq (BSQ) equation.
The main difference is the amount of initial data needed to define
evolution, say one function at $t=0$ for KdV, or one function and its
derivative w.r.t time for BSQ. This has its analogue for partial difference
equations.

\subsection{Initial data configuration}
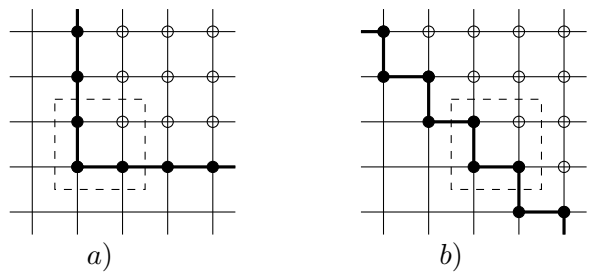
\begin{figure}[h!]
  \centering
\begin{tikzpicture}[scale=0.6]
\foreach \y in {-2,-1,...,2}%
{\draw[thin] (-2.5,\y) -- (2.5,\y);
  \draw[thin] (\y,-2.5) -- (\y,2.5);}
\foreach \y in {-1,...,2}%
 {\node[draw,circle,inner sep=1.5pt,fill] at (-1,\y) {};
\node[draw,circle,inner sep=1.5pt,fill] at (\y,-1) {};}
\foreach \y in {0,...,2}%
\foreach \x in {0,...,2}%
\node[draw,circle,inner sep=1.5pt] at (\y,\x) {};
\draw[very thick] (-1,2.5) -- (-1,-1) -- (2.5,-1);
\draw[dashed] (-1.5,0.5) -- (0.5,0.5) -- (0.5,-1.5)
-- (-1.5,-1.5) -- (-1.5,0.5);
\draw[color=black] (-0.5,-3) node {$a)$};
\end{tikzpicture}\hspace{1.5cm}
\begin{tikzpicture}[scale=0.6]
\foreach \y in {-2,-1,...,2}%
{\draw[thin] (-2.5,\y) -- (2.5,\y);
  \draw[thin] (\y,-2.5) -- (\y,2.5);}
\foreach \x in {-2,...,2}%
\node[draw,circle,inner sep=1.5pt,fill] at (\x,-\x) {};
\foreach \x in {-2,...,1}%
\node[draw,circle,inner sep=1.5pt,fill] at (\x,-\x-1) {};
  \foreach \x in {-2,...,2}%
  {\foreach \y in {-2,...,-\x}%
    \node[draw,circle,inner sep=1.5pt] at (-\y,-\x) {};}
  \foreach \x in {-2,...,1}%
{\draw[very thick] (\x,-\x) -- (\x,-1-\x);
\draw[very thick] (\x,-\x-1) -- (\x+1,-1-\x);}
\draw[dashed] (-0.5,0.5) -- (1.5,0.5) -- (1.5,-1.5)
-- (-0.5,-1.5) -- (-0.5,0.5);
\draw[very thick] (-2,2) -- (-2.5,2);
\draw[very thick] (2,-2) -- (2,-2.5);
\draw[color=black] (-0.5,-3) node {$b)$};
\end{tikzpicture}
\caption{Two ways to give initial data for an equation defined on the
  elementary $2\times 2$ quadrilateral. Here a) gives the corner
  configuration and b) the staircase configuration. The initial data
  is given on the vertices marked by solid black circles and the
  values on vertices marked by open circles are to be computed. The
  points involved in the first step of computations are bounded by the
  dashed line.\label{FT:1}}
\end{figure}

For equations defined on the $\mathbb Z^2$ lattice there are several
possibilities. If the equation is defined on the elementary
quadrilateral of the lattice, such as the discrete versions of the KdV
equation, we need initial data on a line, for example on a corner or
on a staircase as illustrated in Figure \ref{FT:1}. (For more exotic
initial data see \cite{HMW}.)

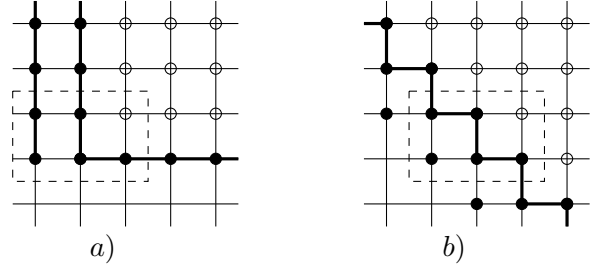
\begin{figure}
  \centering
\begin{tikzpicture}[scale=0.6]
\foreach \y in {-2,-1,...,2}%
{\draw[thin] (-2.5,\y) -- (2.5,\y);
  \draw[thin] (\y,-2.5) -- (\y,2.5);}
\foreach \y in {-1,...,2}%
 {\node[draw,circle,inner sep=1.5pt,fill] at (-1,\y) {};
\node[draw,circle,inner sep=1.5pt,fill] at (-2,\y) {};
\node[draw,circle,inner sep=1.5pt,fill] at (\y,-1) {};}
\foreach \y in {0,...,2}%
\foreach \x in {0,...,2}%
\node[draw,circle,inner sep=1.5pt] at (\y,\x) {};
\draw[very thick] (-1,2.5) -- (-1,-1) -- (2.5,-1);
\draw[very thick] (-2,2.5) -- (-2,-1);
\draw[dashed] (-2.5,0.5) -- (0.5,0.5) -- (0.5,-1.5)
-- (-2.5,-1.5) -- (-2.5,0.5);
\draw[color=black] (-0.5,-3) node {$a)$};
\end{tikzpicture}\hspace{1.5cm}
\begin{tikzpicture}[scale=0.6]
\foreach \y in {-2,-1,...,2}%
{\draw[thin] (-2.5,\y) -- (2.5,\y);
  \draw[thin] (\y,-2.5) -- (\y,2.5);}
\foreach \x in {-2,...,2}%
\node[draw,circle,inner sep=1.5pt,fill] at (\x,-\x) {};
\foreach \x in {-2,...,1}%
\node[draw,circle,inner sep=1.5pt,fill] at (\x,-\x-1) {};
\foreach \x in {-2,...,0}%
\node[draw,circle,inner sep=1.5pt,fill] at (\x,-\x-2) {};
  \foreach \x in {-2,...,2}%
  {\foreach \y in {-2,...,-\x}%
    \node[draw,circle,inner sep=1.5pt] at (-\y,-\x) {};}
  \foreach \x in {-2,...,1}%
{\draw[very thick] (\x,-\x) -- (\x,-1-\x);
\draw[very thick] (\x,-\x-1) -- (\x+1,-1-\x);}
\draw[dashed] (-1.5,0.5) -- (1.5,0.5) -- (1.5,-1.5)
-- (-1.5,-1.5) -- (-1.5,0.5);
\draw[very thick] (-2,2) -- (-2.5,2);
\draw[very thick] (2,-2) -- (2,-2.5);
\draw[color=black] (-0.5,-3) node {$b)$};
\end{tikzpicture}
\caption{Initial data required for equations defined on a $2\times 3$
  stencil, a) for corner and b) for staircase. In both cases, one can
  say that now the initial data needs to be given on ``$1.5$
  lines''. \label{FT:2}}
\end{figure}

Some equations are defined on a $2\times 3$ stencil, for example the
discrete KdV equation in Hirota bilinear form. Two possible ways to
give initial data are shown in Figure \ref{FT:2}.

\begin{figure}[h]
\centering\begin{tikzpicture}[scale=0.6]
\foreach \y in {-3,...,2}%
{\draw[thin] (-3.5,\y) -- (2.5,\y);
  \draw[thin] (\y,-3.5) -- (\y,2.5);}
\foreach \y in {-1,...,2}%
 {\node[draw,circle,inner sep=1.5pt,fill] at (-1,\y) {};
\node[draw,circle,inner sep=1.5pt,fill] at (-2,\y) {};
\node[draw,circle,inner sep=1.5pt,fill] at (\y,-1) {};
\node[draw,circle,inner sep=1.5pt,fill] at (\y,-2) {};}
\node[draw,circle,inner sep=1.5pt,fill] at (-2,-2) {};
\foreach \y in {0,...,2}%
\foreach \x in {0,...,2}%
\node[draw,circle,inner sep=1.5pt] at (\y,\x) {};
\draw[very thick] (-1,2.5) -- (-1,-1) -- (2.5,-1);
\draw[very thick] (-2,2.5) -- (-2,-2) -- (2.5,-2);
\draw[dashed] (-2.5,0.5) -- (0.5,0.5) -- (0.5,-2.5)
-- (-2.5,-2.5) -- (-2.5,0.5);
\draw[color=black] (-0.5,-4) node {$a)$};
\end{tikzpicture}\hspace{1.5cm}
\begin{tikzpicture}[scale=0.6]
\foreach \y in {-2,-1,...,3}%
{\draw[thin] (-1.5,\y) -- (4.5,\y);
  \draw[thin] (1+\y,-2.5) -- (1+\y,3.5);}
\foreach \x in {-1,...,2}%
\node[draw,circle,inner sep=1.5pt,fill] at (\x,-\x) {};
\foreach \x in {-1,...,1}%
\node[draw,circle,inner sep=1.5pt,fill] at (\x,-\x-1) {};
\foreach \x in {-1,...,3}%
\node[draw,circle,inner sep=1.5pt,fill] at (\x,1-\x) {};
\foreach \x in {-1,...,4}%
\node[draw,circle,inner sep=1.5pt,fill] at (\x,2-\x) {};
  \foreach \x in {-2,...,3}%
  {\foreach \y in {-3,...,-\x}%
    \node[draw,circle,inner sep=1.5pt] at (1-\y,1-\x) {};}
  \foreach \x in {-1,...,1}%
{\draw[very thick] (\x,-\x) -- (\x,-1-\x);
\draw[very thick] (\x,-\x-1) -- (\x+1,-1-\x);}
\foreach \x in {-1,...,3}%
{\draw[very thick] (\x,2-\x) -- (\x,1-\x);
\draw[very thick] (\x,-\x+1) -- (\x+1,+1-\x);}
\draw[dashed] (-0.5,1.5) -- (2.5,1.5) -- (2.5,-1.5)
-- (-0.5,-1.5) -- (-0.5,1.5);
\draw[very thick] (-1,3) -- (-1.5,3);
\draw[very thick] (-1,1) -- (-1.5,1);
\draw[very thick] (2,-2) -- (2,-2.5);
\draw[very thick] (4,-2) -- (4,-2.5);
\draw[color=black] (-0.5,-3) node {$b)$};
\end{tikzpicture}
\caption{Initial data required for equations defined on a $3\times 3$
  stencil, a) for corner, b) for staircase. In this case, the initial
  data needs to be given on 2 infinite lines. For the staircase the
  two lines can be defined by $0\le n+m \le 3$. \label{FT:3}}
\end{figure}
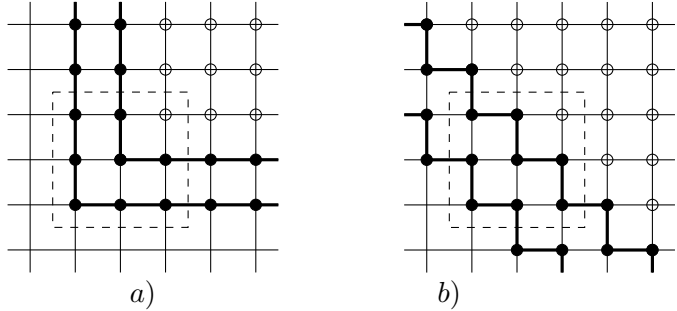

In this paper, we study one component equations defined on a $3\times
3$ stencil on the $\mathbb Z^2$ lattice. For them two lines of initial
data is necessary, as seen in Figure \ref{FT:3}.

\subsection{The equations under study\label{S:eqs}}
\begin{figure}
\centering
\begin{tikzpicture}[scale=0.8]
\foreach \y in {-2,-1,...,2}%
\draw[thin] (-2.5,\y) -- (2.5,\y);
 \node[draw,circle,inner sep=1.5pt,fill] at (0,0) {};
 \node[draw,circle,inner sep=1.5pt,fill] at (-1,0) {};
 \node[draw,circle,inner sep=1.5pt,fill] at (1,0) {};
 \node[draw,circle,inner sep=1.5pt,fill] at (0,0) {};
 \node[draw,circle,inner sep=1.5pt,fill] at (1,1) {};
 \node[draw,circle,inner sep=1.5pt,fill] at (0,-1) {};
 \node[draw,circle,inner sep=1.5pt,fill] at (0,1) {};
 \node[draw,circle,inner sep=1.5pt,fill] at (-1,-1) {};
 \draw[thin,dashed] (1,1) -- (-1,-1);
 \draw[thin,dashed] (0,-1) -- (0,1);
\foreach \x in {-2,-1,...,2}%
 \draw[thin] (\x,-2.5) -- (\x,2.5);
\node at (-0.5,-3) {a)};
\end{tikzpicture}\hspace{1.5cm}
\begin{tikzpicture}[scale=0.8]
\foreach \y in {-2,-1,...,2}%
\draw[thin] (-2.5,\y) -- (2.5,\y);
 \node[draw,circle,inner sep=1.5pt,fill] at (0,0) {};
 \node[draw,circle,inner sep=1.5pt,fill] at (0,-1) {};
 \node[draw,circle,inner sep=1.5pt,fill] at (0,1) {};
 \node[draw,circle,inner sep=1.5pt,fill] at (1,1) {};
 \node[draw,circle,inner sep=1.5pt,fill] at (-1,-1) {};
 \node[draw,circle,inner sep=1.5pt,fill] at (-1,1) {};
 \node[draw,circle,inner sep=1.5pt,fill] at (1,-1) {};
 \draw[thin,dashed] (1,1) -- (-1,-1);
 \draw[thin,dashed] (1,-1) -- (-1,1);
\foreach \x in {-2,-1,...,2}%
 \draw[thin] (\x,-2.5) -- (\x,2.5);
\node at (-0.5,-3) {b)};
\end{tikzpicture}
\caption{a): Graph of the integrable 4-term 7-point equation
  \eqref{eq:HB7int} b): Graph of the non-integrable 4-term 7-point
  equation \eqref{eq:HB7non}. The (0,0) point at is the lower left
  black disk.
 \label{FT:4}}
\end{figure}
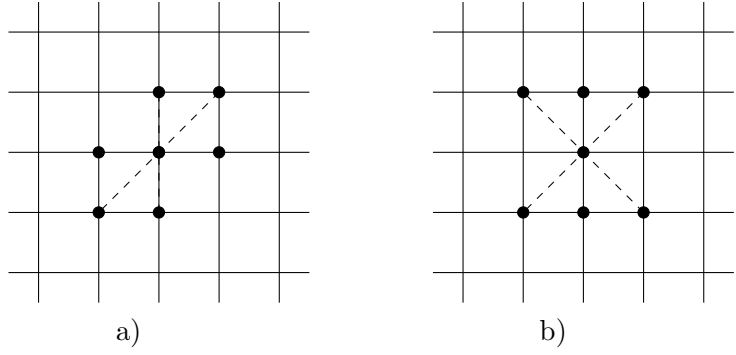

The first pair of equations are in the Hirota bilinear form. We use
the short-hand notation of writing only the shift with respect to
$(n,m)$ thus $f_{n+1,m}\equiv f_{1,0}$ etc. The two equations considered are:
\begin{itemize}
\item Four-term seven point integrable equation in Hirota bilinear form
  (see Figure \ref{FT:4}a))
\begin{equation}\label{eq:HB7int}
2f_{2,2}f_{0,0}+2f_{1,2}f_{1,0}-f_{2,1}f_{0,1}-3f_{1,1}^2=0.
\end{equation}

\item Four-term seven point non-integrable equation in Hirota bilinear
  form (see Figure \ref{FT:4}b))
\begin{equation}\label{eq:HB7non}
  12\,f_{2,2}f_{0,0} - 3\,f_{2,0}f_{0,2} + 16\,f_{1,2}f_{1,0}
  - 25\,f_{1,1}^2=0.
\end{equation}
\end{itemize}
The numerical values of the parameters in \eqref{eq:HB7int} and
\eqref{eq:HB7non} have no effect on integrability; they have been
chosen for convenience of integrability analysis using the
three-soliton-condition (3SC), done in the Appendix.
There are also integrable three-term Hirota bilinear equations
depending on 5 points of the 9 point stencil. They are related to the
Toda lattice and we will not discuss them here.

The second set of equations consist of one-component equations of
Boussinesq type. (For a comprehensive review see \cite{BSQREV}.) Their
original three-component forms are defined on the elementary
quadrilateral and on its boundaries and are known to be integrable by
Consistency-Around-the-Cube method \cite{Hie2011}. By eliminating two
variables one obtains these one-component forms on a larger
stencil. For them no independent proof of integrability has been given
so far. In order to get a non-integrable version of these
equations we will add a multiplier ($\neq 1$) to the $x_{0,0}$ terms.
In the following $P,Q$ are the lattice parameters.

\begin{itemize}
\item The regular lattice Boussinesq equation in one-component form 
\begin{align}
(P-Q)\left(\frac1{x_{2,0}-x_{1,1}}-\frac1{x_{1,1}-x_{0,2}}\right)
+b_0(x_{0,1}-x_{1,0}+x_{2,1} -x_{1,2})&\nonumber  \\
-(x_{2,2}-x_{0,1})(x_{2,1}-x_{1,2})
-(x_{0,0}-x_{2,1})(x_{1,0}-x_{0,1})&=0. \label{eq:B2-x-only}
\end{align}
This was first given in \cite {NPCQ92},
except for the parameter $b_0$ found in \cite{Hie2011}.
\footnote{ It turns out that the $b_0$ term in \eqref{eq:B2-x-only}
  can be eliminated by the transformation $x_{n,m}\to
  x_{n,m}+\tfrac13b_0(n+m)$. However, the parameter $b_0$ cannot be
  eliminated from the three-component form and in fact has effects on
  the solutions\cite{HZtaxo}.}

\item The modified lattice Boussinesq equation in one-component form 
 was first given in \cite {NPCQ92}, we use its reversed form with
 $x_{2,2}$ in the numerator
\begin{equation}
 \left(\frac{P\, x_{1,1}-Q\, x_{2,0}}{x_{2,0}-x_{1,1}}\right)
  \frac{x_{1,0}}{x_{2,1}}
-\left(\frac{P\, x_{0,2}-Q\, x_{1,1}}{x_{1,1}-x_{0,2}}\right)
\frac{x_{0,1}}{x_{1,2}}
=\frac{x_{2,2}}{x_{1,2}}-\frac{x_{2,2}}{x_{2,1}}-\frac{x_{1,0}}{x_{0,0}}+
\frac{x_{0,1}}{x_{0,0}}.\label{eq:A2-x-only}
\end{equation}

\item The Schwarzian Boussinesq equation in one-component form 
\begin{align}
 &\frac{(x_{2,2}-x_{1,2})(x_{0,2}-x_{1,1})(x_{0,1}-x_{0,0})}
{(x_{2,2}-x_{2,1})(x_{1,1}-x_{2,0})(x_{1,0}-x_{0,0})}=\nonumber\\
&\hskip -1.7cm\frac{(x_{1,1}-x_{0,2})(b_0\,x_{0,1}+b_1)
+(x_{1,2}-x_{0,2})(x_{0,1}-x_{1,1})\,P-
(x_{1,2}-x_{1,1})(x_{0,1}-x_{0,2})\,Q}
{(x_{2,0}-x_{1,1})(b_0\,x_{1,0}+b_1)
+(x_{2,1}-x_{1,1})(x_{1,0}-x_{2,0})\,P-
(x_{2,1}-x_{2,0})(x_{1,0}-x_{1,1})\,Q}.\label{eq:C34-x-only}
\end{align}
This was first given in \cite{N-AAIS-1997} except for the parameters
$b_0,\,b_1$ found in \cite{Hie2011}.
\end{itemize}

\section{Degree growth computations\label{C:comp}}
\subsection{Limits of analytical computations}
For precise degree growth analysis one could in principle use analytical
computations. In order to get an idea of its feasibility, let us
consider the integrable equation \eqref{eq:A2-x-only}.  For
computations we write equation \eqref{eq:A2-x-only} for \CP2 with a
pair of polynomial maps. Defining polynomials $y,a$ by
\[
x_{n,m}=\frac{y_{n,m}}{a_{n,m}},
\]
we get the maps
\begin{eqnarray*}
y_{n,m}&=&[
(a_{n - 2,m}\,y_{n - 1,m - 1}\,Q - a_{n - 1,m - 1}\,y_{n - 2,m}\,P)\times\\
&&(a_{n - 1,m - 1}\,y_{n,m - 2} - a_{n,m - 2}\,y_{n - 1,m - 1})\,
a_{n - 1,m - 2}\,a_{n - 1,m}\,y_{n - 2,m - 1}\,y_{n,m - 1}\\
&&\hskip -0.3cm - (a_{n - 1,m - 1}\,y_{n,m - 2}\,Q - a_{n,m - 2}\,y_{n - 1,m - 1}\,P)
\times\\
&&(a_{n - 2,m}\,y_{n - 1,m - 1} - a_{n - 1,m - 1}\,y_{n - 2,m})
a_{n - 2,m - 1}\,a_{n,m - 1}\,y_{n - 1,m - 2}\,y_{n - 1,m}]y_{n - 2,m - 2}\\
&&\hskip -0.3cm +
(a_{n - 1,m - 1}\,y_{n,m - 2} - a_{n,m - 2}\,y_{n - 1,m - 1})
\,(a_{n - 2,m}\,y_{n - 1,m - 1} - a_{n - 1,m - 1}\,y_{n - 2,m})\times\\
&&(a_{n - 2,m - 1}\,y_{n - 1,m - 2} - a_{n - 1,m - 2}\,y_{n - 2,m - 1})\,
a_{n - 2,m - 2}\,y_{n - 1,m}\,y_{n,m - 1},\\
a_{n,m}&=&(a_{n - 2,m}\,y_{n - 1,m - 1} - a_{n - 1,m - 1}\,y_{n - 2,m})\,
(a_{n - 1,m - 1}\,y_{n,m - 2} - a_{n,m - 2}\,y_{n - 1,m - 1})\times
\\&&(a_{n - 1,m}\,y_{n,m - 1} - a_{n,m - 1}\,y_{n - 1,m})\,
a_{n - 2,m - 1}\,a_{n - 1,m - 2}\,y_{n - 2,m - 2}.
\end{eqnarray*}
We take the staircase configuration and as initial values all
$x_{n,m}$ for $n+m=0,1,2,3$ are free parameters. In practice, we take
$y_{n,m}=f_{n,m},\,a_{n,m}=1$, when $n+m=0,1,2,3$.  The first computed
values at $n+m=4$, for example $y_{2,2},\,a_{2,2}$, will respectively
be of degrees 5 and 4 in $f$, and will have 16 resp.\ 8 terms,

Continuing with $n+m=5$, we get $y_{3,2}$ and $a_{3,2}$ of degrees 13
and 12 (with 1184 and 528 terms), respectively. But there are
cancellations. We find
\[
\text{GCD}(y_{3,2},a_{3,2})=(f_{3,0}-f_{2,1}) (f_{2,1}-f_{1,2})(f_{2,0}-f_{1,1})f_{2,1},
\]
and consequently after cancellation the final degrees are 9 and 8,
respectively (with 220 and 112 terms). The GCD also clearly indicates
that initial values with ${f_{n,m}-f_{n-1,m+1}=0}$ form singular lines
in the initial value space coordinatized by ${f_{n,m},\,n+m=0,1,2,3}$.

Getting data for points with $n+m=6$ is already very demanding, but
with judicious choices in the order of computations we eventually find
that GCD$(y_{3,3},a_{3,3})$ is a product of various
${f_{n,m}-f_{n-1,m+1}}$ terms and when they are divided out
$y_{3,3},a_{3,3}$ will have 4672 and 2592 terms of degrees 14 and
13, respectively.  Thus with some effort we have found the beginning
part of the degree growth sequence to be (c.f. Section \ref{S:MLB})
\[
1,\,5,\,9,\,14,\dots
\]
It is now clear that with full analytical computations we cannot hope
to get a sufficient number of data points for growth analysis. And for
the more complicated \eqref{eq:C34-x-only} the situation is worse
still.  Thus, we must resort to other techniques.

\subsection{Computational method}
In the following, we will compute numerically the degree growth for
the equations mentioned in Section \ref{S:eqs}. We take the equations
in Hirota bilinear form as test cases, because for them we have an
independent integrability test. Thus, from these equations we get an
indication on how the computed degrees can be used to distinguish
between integrable and non-integrable models. We then go on to compute
degrees for the Boussinesq equations.

As we saw a full analytic computation not feasible, but for longer
growth data even the numerical computations must be streamlined:
\begin{itemize}
  \item We must reduce the number of variables that are tracked for
    determining the degree. It is not necessary to keep all
    initial values as independent variables, instead one can introduce
    one dedicated variable, say $z$, for this purpose and give the
    initial data in terms of that variable, for example:
    \begin{enumerate}
    \item Use the dedicated variable at one point only, say
      $f_{0,0}=z$, all other initial values being random numbers.
    \item Set all initial values as linear functions of the dedicated
      variable $z$: \begin{equation}
        f_{n,m}=\alpha_{n,m}z+\beta_{n,m}, \end{equation} with random integer
      coefficients $\alpha, \beta$. Then we take $y_{n,m}=f_{n,m},\, a_{n,m}=1$.
    \item Set all initial values as rational functions of $z$
      \cite{Viallet2006}
    \begin{equation}
    f_{n,m}=\frac{\alpha_{n,m}z+\beta_{n,m}}{\gamma_{n,m}z+\delta_{n,m}}.
    \end{equation}
 For polynomial computations we set
 $y_{n,m}={\alpha_{n,m}z+\beta_{n,m},\,}
 a_{n,m}={\gamma_{n,m}z+\delta_{n,m}}$.
    \end{enumerate}
  \item During iterations the expressions start to contain huge
    numbers as coefficients and this can be made manageable using
    modular arithmetic with respect to a large prime $p$. It is best
    to use random numbers for all constants (such as those appearing
    in the equation $P,Q,b_j$, as well as those appearing in initial
    conditions $\alpha_{n,m},\beta_{n,m},\dots$). The necessary operations of
    polynomial algebra, including computation of GCD, have been
    implemented for modular arithmetic.  The programming language C++
    provides tools for this purpose through its NTL package. In
    principle, there could be spurious zeroes if by accident some
    number is congruent to 0 mod $p$. In a suspicious situation one
    could redo computations with a different prime.
  \item Initial value geometry. The choices we will use here,
    corner and staircase, were already illustrated in Figures
    \ref{FT:2} and \ref{FT:3}. We do not consider exotic geometries
    such as those used in \cite{HMW}.
\end{itemize}

\section{Degree growth for Hirota bilinear equations\label{C:H}}
Integrability analysis of equations \eqref{eq:HB7int} and
\eqref{eq:HB7non} based on the three-soliton-condition has been done
in the Appendix. In this section we study the degree growth. We
will consider different geometries and distributions of $z$.

\subsection{Only one tracking variable among the initial value}
\paragraph{The integrable case \eqref{eq:HB7int}.}
First we will consider the integrable equation \eqref{eq:HB7int} and
put the tracking variable $z$ in only one initial value.\footnote{This
  method was first used in \cite{HMW} where it was found that for the
  Liouville equation the degrees are then bounded, for the integrable
  KdV equation the degrees grow linearly, while in a non-integrable
  version of KdV the degrees were found to grow asymptotically as
  $4^n$.}

\begin{figure}
\centering   
\begin{tikzpicture}[scale=0.7]   
  \input hmistrone25tikz11s
  \stenhmi{11}{11}
 \draw[dashed] (10.5,10.5) -- (10.5,24);
 \draw[dashed] (10.5,10.5) -- (24,10.5);

\end{tikzpicture}
\caption{Degrees for equation \eqref{eq:HB7int}. The variable $z$ is
  only at $(1,1)$. The corner configuration is within the dashed
  lines. For $n,m>5$ the degrees are
  $\deg_{n,m}=4\min(n,m)-10$.\label{F:H7is11}}
\end{figure}
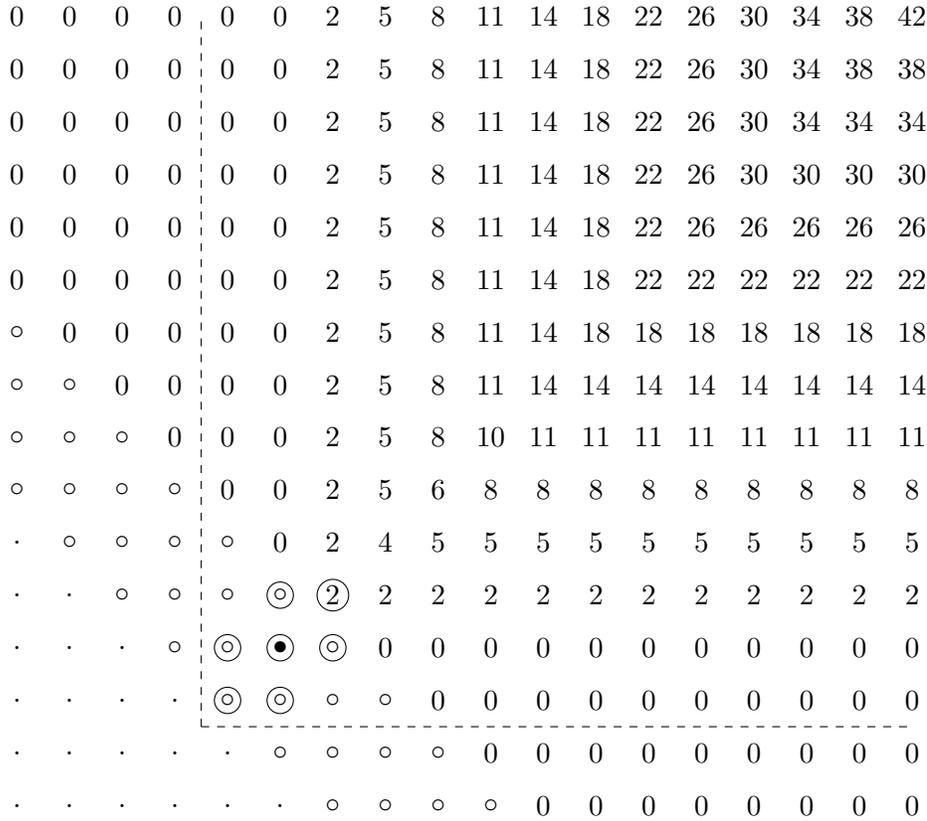
In Figure \ref{F:H7is11}, we have the staircase configuration with the
initial value $z$ at $(1,1)$, it is marked by a black disk. The other
initial values are random numbers and marked by open circles. We have
also displayed the stencil relevant to this equation using larger
circles. The displayed numbers give the degree of the numerator of
$x_{n,m}$, i.e., the degree of $y_{n,m}$ in $z$. For $n,m>5$ the degrees
are given by
\beq
  \deg_{n,m}=4\min(n,m)-10,
\eeq
and therefore grow linearly.

In the same figure, we have indicated the corner configuration with
dashed lines. The degrees are the same, because from staircase of
initial values we can compute numerical values for all points for
which $(n<2,\, m>2)$ or $(n>2, m<2)$, and this way create the initial
values (random numbers) for the  corner situation.

If we set the initial value $z$ at $(0,0)$, the degree growth is slower.
For example, the degree at the first calculated point $(2,2)$ is zero,
because the variable $z$ only appears in the denominator.  For $n,m>2$
the degrees are in that case given by
\beq
\deg_{n,m}=2\min(n,m)-4.
\eeq
In all cases, when there is only one $z$-dependent initial value the
growth is linear for the integrable case.

\paragraph{The non-integrable case \eqref{eq:HB7non}.}


For the non-integrable case \eqref{eq:HB7non} the degrees in Figure
\ref{F:H7ns11} are for the initial value $z$ only at $(1,1)$, in a
staircase configuration. A longer list of values on the diagonal
(dashed line in Figure \ref{F:H7ns11}) is given by
\begin{equation*}
0,\, 1 \vert \, 2,\, 4,\, 8,\, 22,\, 56,\, 151,\, 402,\, 1103,\, 3010,\,
8324,\, 23034,\, 64171,\, 179096, 501810,
\,1408760,
\,\dots
\end{equation*}\begin{figure}
\centering   
    {\small \begin{tikzpicture}[scale=0.7]  
    \input hmnstrone23tikz11slew
     \stenhmn{10}{10}{1.38}
     \draw[dotted] (13.8,10) -- (22*1.38,22);
    \end{tikzpicture}}
\caption{Non-integrable equation \eqref{eq:HB7non} with initial single
  $z$ at point $(1,1)$.
  \label{F:H7ns11}}
\end{figure}
This is approximately $ 0.726\times 2.813^{n}$. The first few columns
in the figure show regular growth but is still unlikely that the
numbers can be given by some formula, especially since the equation
itself is asymmetric.

\subsection{All initial values contain $z$}
Next we consider cases in which all initial values depend on $z$,
either linearly, i.e., $\alpha_{n,m}z+\beta_{n,m}$ or rationally
$(\alpha_{n,m}z+\beta_{n,m})/(\gamma_{n,m}z+\delta_{n,m})$, where $\alpha_{n,m},\,\beta_{n,m},\,
\gamma_{n,m},\, \delta_{n,m}$ are some random numbers. We will mostly use the
staircase configuration extending from upper left to lower right.
As indicated in Figure \ref{FT:3}, the initial values will be given on
points for which $0\le n+m \le 4$. Due to translational invariance
along the staircase the degrees only depend on $n+m$ and therefore the
results can be given by one sequence of values.

\paragraph{Integrable/linear.}
In the staircase configuration the degrees for \eqref{eq:HB7int}
are given by
\beq\label{eq:yht2}
\deg_{k=n+m>3}=k^2-7k+14.
\eeq
We also computed the degrees without
canceling common factors, and got (for $k>3$)
\begin{equation*}
  2, 4, 10, 28, 81, 237, 697,
2053, 6050, 17832, 52562, 154936, 456705, 1346233, 3968305, \dots,
\end{equation*}
leading to growth rate $2.94771^k$, showing once more the essential
influence of cancellations.


For the corner configuration and with linear initial data the degrees
of the numerator follow the rule \beq\label{eq:yht1} \deg_{n,m} = 2nm
- 3(n + m) + 6.  \eeq On the diagonal where $m=n$, we have $k=2n$ and
we can compare staircase and corner degrees: $\deg_{str}=4n^2-14n+14$
versus $\deg_{cor}=2n^2-6n+6$, i.e., in the staircase the growth is
about twice as fast. This is because in the corner case the degrees on
the corner boundaries (cf. Figure \ref{FT:3} a)) are those of the
initial values (i.e. $=1$) while for the staircase the degrees at the
corresponding points are computed.

\paragraph{Integrable/rational.}

In Figure \ref{F:H7icorrat}, we have rational initial data for the
corner configuration. In the region above and to the right of the
dashed line the degrees are given by
\beq
\deg_{n,m} = 6mn - 13(n + m)+4\max(n, m) + 2\delta_{n,m} + 32.
\eeq
Asymptotically these are three times bigger than in the linear case
\eqref{eq:yht1}.
\begin{figure}[t]
\centering
{\scriptsize
 \begin{tikzpicture}[scale=0.72]  
  \input hmicorrat20tikzcut
 \draw[dashed] (5.5,5.5) -- (5.5,15.5);
 \draw[dashed] (5.5,5.5) -- (19,5.5);
\end{tikzpicture}}
 \caption{Degrees for equation \eqref{eq:HB7int} with rational initial
   data.\label{F:H7icorrat}}
\end{figure}

For the staircase we have the degrees starting at $k=4$ as
$7,13,26,43,62,91,122$ and then for $k>10$,
\beq\deg_{k=n+m}=3k^2-26k+84,\eeq again about three times bigger than
in \eqref{eq:yht1}.

\paragraph{Non-integrable/linear}
For the staircase configuration we have the degrees for
$k:=n+m$, starting with $k=0$, as
\begin{align*} 1, 1, 1, 1,\vert 2, 3, 7, 12, 22,
36, 61, 101,& 174, 295, 508, 864, 1478, 2513, 4289, 7303, 12463,\\& 21241,
36237, 61771, 105346, 179593, 306252, \dots.
\end{align*}
This is
approximately $\propto 1.705^k$ or on the $n=m$ diagonal as $\propto
2.908^n=1.705^{2n}$.

\paragraph{Non-integrable/rational}
For the staircase the degree sequence is
\begin{align*}1, 1, 1, 1, \vert 7, 11, 24,
38, 64, 102,& 176, 294, 514, 870, 1498, 2539, 4341, 7376, 12600,\\& 21456,
36631, 62419, 106488, 181496, 309541,
\dots. \end{align*}
The is about the same as for linear initial data

\paragraph{Summary.} Depending on the initial data from which the
degrees are computed, we have slightly different growth rates. In the
integrable case of \eqref{eq:HB7int} a single $z$ dependent initial
data point gives linear growth, while linear and rational initial data
give quadratic growth.  For the chosen example of non-integrable
equation in Hirota bilinear form \eqref{eq:HB7non} the growth is
always exponential, about $2.8^n$ or $2.9^n$.  Thus, we see that the
computationally simplest case of only one initial $z$-dependent point
is enough to differentiate between integrable and non-integrable
equations.

\section{Degree growth of lattice Boussinesq equations\label{C:B}}
We now turn to the lattice Boussinesq equations \eqref{eq:B2-x-only},
\eqref{eq:A2-x-only}, and \eqref{eq:C34-x-only}. These equations
involve all points of the $3\times3$ stencil. The non-integrable
versions are obtained by changing the coefficient of the $x_{0,0}$
term.

From the results for Hirota equations \eqref{eq:HB7int} and
\eqref{eq:HB7non} we have observed first of all that a single
non-numeric initial value is enough to differentiate between
integrable and non-integrable equations. Furthermore, we found that
even for equations defined on a larger stencil, the integrable case
with linear and rational $z$ dependence have quadratic degree growth,
while for the non-integrable case the growth is typically $2.8^n$. One
may expect similar overall results for the Boussinesq
equations since they are integrable.

However we may expect differences in the details. One difference is
due to the observations that Boussinesq equations are in many ways
associated with threefold symmetry, for example the solutions often
involve cubic roots of unity. In the following, this manifests itself
in the degree sequences where we need an indicator function for
divisibility by 3, which we define as
\[
\mathcal{D}_n(m)=\left\{\begin{array}{c l}
    1 & \text{ when } n|m,\\
    0 & \text{ otherwise.}
  \end{array}\right.
  \]
  
\subsection{Regular lattice Boussinesq equation \eqref{eq:B2-x-only}}


For numerical computations we use $P-Q=3,\,b_0=1$. The value of $b_0$
seems to have no effect. Unless mentioned otherwise, we only consider
the staircase configuration.

\begin{figure}
\centering
   {\scriptsize
\begin{tikzpicture}[scale=0.6]   
      \input 310corone17tikz13000cc
   \draw[dashed] (5.5,16) -- (5.5,10) -- (10,5.5) -- (16,5.5);
   \draw[dashed] (6.5,16) -- (6.5,12) -- (12,6.5) -- (16,6.5);
   \stenbsq{0}{0}
\end{tikzpicture}}
\caption{Degrees for Boussinesq equation \eqref{eq:B2-x-only} with one
  $z$ at (0,0). The degrees are given by
  $\min(n-1,m-1,\lfloor(n+m-1)/3\rfloor)$. As an example, the dashed
  lines border the area with degree 5.
  \label{F:B310z}}
\end{figure}
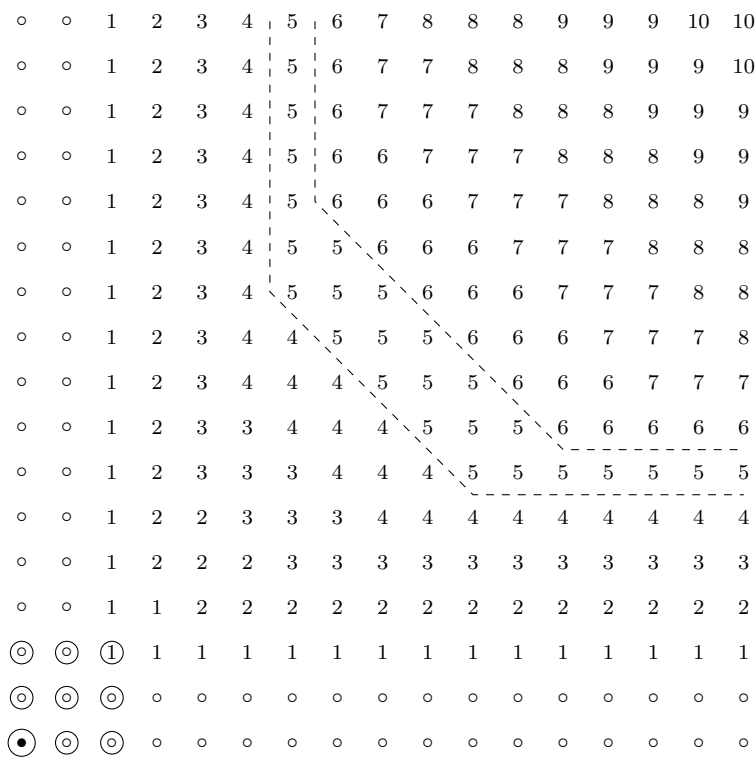

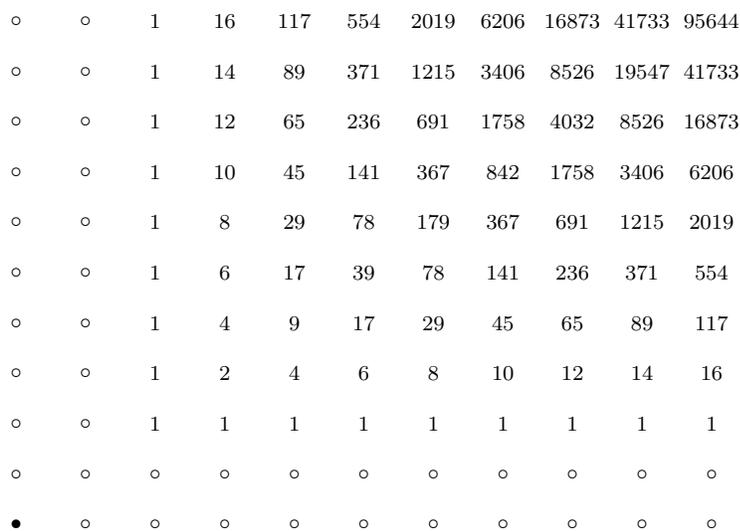
\begin{figure}
\centering 
   {\scriptsize
\begin{tikzpicture}[scale=0.67]   
      \input 310corone11tikz13100lew
   \end{tikzpicture}}
\caption{Degrees for a non-integrable case. First difference with
  respect to the integrable case displayed in Figure \ref{F:B310z} is
  at point $(3,3)$. \label{F:B310znon}}
\end{figure}

{\bf $\bullet$ Single $z$ in the initial values.} If initial value has
a single $z$ at $(0,0)$ we get for the integrable case degrees for the
corner configuration as given in Figure
\ref{F:B310z},\footnote{$\lfloor x \rfloor=$ ``floor'' of $x$ = ignore
  decimals of $x$.}  \beq\label{eq:B310z}
\deg_{n,m}=\min(n-1,m-1,\lfloor(n+m-1)/3\rfloor) \eeq The same degrees
are obtained for the staircase configuration.
In Figure \ref{F:B310znon}, we have degrees for a case which is
non-integrable due to a different coefficient for $x_{0,0}$.



{\bf $\bullet$ Linear initial values:}

In the integrable case
the degrees in the staircase configuration are
\begin{align*}
  1, 1, 1, 1\,|\, 4, 7, 11, 16, 22, 29, 37, 46, 56, 67, 79, 92,
  & 106, 121, 137, 154,
  172, 191, 211,\\& 232, 254, 277, 301, ...
\end{align*}
  Here and in the following, the first 4 degrees are initial values at
  $(0,0), \,(0,1), \,(1,1),\, (1,2)$ The above sequence is given by \beq
  \deg_{k=n+m>3}=\tfrac12(k^2-3k+4).  \eeq

\vskip 0.3cm

In the non-integrable case with ``$2\,x_{0,0}$'' instead of ``$x_{0,0}$''
we have instead
\begin{align*}
1, 1, 1, 1\,|\, 4, 7, 13, 24, 47, 93, 180, 353, 695,& 1358, 2655, 5206,
10192, 19942,\\& 39048, 76447, 149634, 292944, 573525,\dots
\end{align*}
with growth rate $0.112\cdot 1.958^k$.  First difference with respect
to the integrable case is at $(3,3)$: 13 vs. 11.

{\bf $\bullet$ Rational initial values}

Integrable case we have degrees 
\begin{align*}
 1, 1, 1, 1\,|\, 9, 16, 26, 41, 55, 73, 97, 118, 144, 177,& 205, 239,
 281, 316, 358, 409,\\& 451, 501, 561, 610, 668, 737, 793, \dots
\end{align*}
This is fitted with
\beq\label{eq:degratb}
\deg_{k=n+m>3}=\tfrac13[4k^2-13(k-1)+(k-2)\mathcal{D}_3(k-1)-\mathcal{D}_3(k)].
\eeq
Note the period-3 components.

\vskip 0.3cm

For the non-integrable case  with ``$2\,x_{0,0}$'' instead of ``$x_{0,0}$'' we get
\begin{align*}
 1, 1, 1, 1\,|\, 9, 17, 33, 63, 123,& 243, 473, 927, 1823, 3567, 6977,
 13675, 26777,\\&  52403, 102599, 200863, 393179, 769723, 1506935, \dots
\end{align*}
 The growth is about $0.29\cdot 1.958^k$.  With the $x_{0,0}$ term
 replaced by ``$0\,x_{0,0}$'' we get slightly smaller degrees
\begin{align*}
1, 1, 1, 1\,|\, 8, 15, 29, 55, 108,& 213, 414, 813, 1598, 3125, 6115,
11985, 23464,\\& 45923, 89915, 176023, 344559, 674551, 1320600,\dots
\end{align*}
with about the same growth: $0.26\cdot 1.958^k$. Although the degrees
start slower in the last case they exceed the integrable case already at
$k=6$.

\subsection{Modified lattice Boussinesq equation \eqref{eq:A2-x-only}
  \label{S:MLB}}
In general the degrees are close to those of the regular Boussinesq
equation, sometimes even the same. In computations we use $p=1,q=3$.

{\bf $\bullet$ Single $z$ in the initial values.} The degrees are the
same as for the regular Boussinesq equation, given in
\eqref{eq:B310z} and Figure \ref{F:B310z}.

{\bf $\bullet$ Linear initial values:}

For the integrable case 
we get
\begin{align*}
1, 1, 1, 1,\, |\, 5, 9, 14, 21, 29, 38, 49, 61, 74, 89,\  & 105, 122 ,141,
161, 182, 205,\\& 229, 254, 281, 309, 338, 369, 401, ...
\end{align*}

This is fitted with
\beq
\deg_{k=n+m>3}=\tfrac13[2k^2-6k+7-\mathcal{D}_3(n)].
\eeq

\vskip 0.3cm

For the non-integrable case
\begin{align*}
1, 1, 1, 1,\, |\, 5, 10, 21, 49, 112, 255, 582,& 1329, 3035, 6930, 15824, 36134,\\& 82511,
188411, 430231, 982420, 2243327, ...
\end{align*}
The asymptotic degree growth is approximately $0.151\cdot 2.283^n$.

{\bf $\bullet$ Rational initial values}

For the integrable case 
\begin{align*}
1, 1, 1, 1,\, |\, 9, 17, 27, 41, 57, 75,  97, 121,\,& 147, 177, 209, 243,
281, 321, 363,  \\& 409, 457, 507, 561, 617, 675, 737, 801, ...
\end{align*}
This is fitted with
\beq\label{eq:degrata}
\deg_{k=n+m>3}=\tfrac13[4k^2-12k+11-2\mathcal{D}_3(k)].
\eeq
This bears some similarity to \eqref{eq:degratb}. Indeed we have
\beq
\deg^{\eqref{eq:degratb}}-\deg^{\eqref{eq:degrata}}=(\mathcal{D}_3(n-1)-1)
\lfloor(n-1)/3\rfloor
\eeq
which means that every third degree value is the same.

\vskip 0.3cm

The non-integrable case gives 
\begin{align*}
1, 1, 1, 1,\, |\,  9, 18, 38, 90, 206, 469, 1071,\ & 2446, 5586, 12755,\\& 29125, 66507,
151867, 346783, 791869, ...
\end{align*}
Now the growth is approximately $0.279\cdot 2.283^n$.

\subsection{Schwarzian lattice Boussinesq equation \eqref{eq:C34-x-only}}
It turns out that in the integrable case the degrees are the same as
for the modified Boussinesq equation. Degrees for the
non-integrable cases are different, however.

\subsubsection{A generalization}
In \cite{ZZN12} a generalization for the Schwarzian lattice Boussinesq
equation was given in the form
\begin{equation}\label{eq:interBSQ-e}
\frac{{\mathcal Q}_{p,q}(x_{0,1},x_{1,1},x_{0,2},x_{1,2})}
{{\mathcal Q}_{p,q}(x_{1,0},x_{2,0},x_{1,1},x_{2,1})}
= \frac{(Q_a x_{0,0}-Q_bx_{0,1})\,(P_a x_{1,2}-P_b
x_{2,2})\,(Q_aP_bx_{1,1}-P_a Q_bx_{0,2})}
{(P_a x_{0,0}-P_bx_{1,0})\,(Q_a x_{2,1}-Q_b
x_{2,2})\,(Q_aP_bx_{2,0}-P_a Q_bx_{1,1})},
\end{equation}
in which ${\mathcal
  Q}_{p,q}(x_{0,1},x_{1,1},x_{0,2},x_{1,2})$ and
${\mathcal Q}_{p,q}(x_{1,0},x_{2,0},x_{1,1},x_{2,1})$
can be obtained from
\begin{eqnarray}\label{eq:Q}
  {\mathcal Q}_{p,q}(x_{0,0},x_{1,0},x_{0,1},x_{1,1})&:=&
  P_a P_b (x_{0,0} x_{0,1}+x_{1,0}x_{1,1})\\&&
- Q_a Q_b(x_{0,0}x_{1,0}+x_{0,1}x_{1,1})-G(p,q)  (x_{0,1}x_{1,0}+ x_{0,0} x_{1,1})\ ,\nn
\end{eqnarray}
by $m$ and $n$ shifts, respectively. Furthermore, the parameters $P,
Q, G$ are given by \beq
P_a^2=g(p)-g(a),\,P_b^2=g(p)-g(b),\,Q_a^2=g(q)-g(a),\,Q_b^2=g(q)-g(b),\,
G(p,q)=g(p)-g(q), \eeq where
$g(x)=x^3-\alpha_2x^2+\alpha_1x$. Actually the form of the function
$g$ is irrelevant, because only parameters $P,Q,G$ enter in the
equation, and due to their additive definition they can be considered
free, except for the following constraints \beq
P_a^2-P_b^2=Q_a^2-Q_b^2,\quad G(p,q)=P_a^2-Q_a^2.  \eeq
In numerical computations we took
$P_a=40,\,P_b=32,\,Q_a=24,\,Q_b=7$. We found that this equation has
the same degrees as the  standard Schwarzian lattice Boussinesq equation.

If $P_a^2=P_b^2$, $Q_a^2=Q_b^2$ \eqref{eq:interBSQ-e} reduces to
\eqref{eq:C34-x-only} with $b_0=b_1=0$. 

It was surmised in \cite{ZZN12} that from \eqref{eq:interBSQ-e} one
can obtain the other one-component lattice Boussinesq equations by
suitable transformations and limits, but no rigorous proofs have been
presented. If such connections exist they are not simple. For example
\eqref{eq:interBSQ-e} is homogeneous and scale invariant while the
$b_1$ term in \eqref{eq:C34-x-only} breaks that. Furthermore the $b_i$
terms in \eqref{eq:B2-x-only} and \eqref{eq:C34-x-only} arise from the
$\alpha_i$ terms in $g(x)$ but in \eqref{eq:interBSQ-e} the $\alpha_i$
terms are entirely hidden in $P,Q,G$.

\section{Discussion}
The lattice Boussinesq equations are usually given as three component
equations residing on a single lattice plaquette and on its
boundaries.  By eliminating certain variables in favor of others one
can obtain \cite{BSQREV} several one component representation of the
three kinds of Boussinesq equations. It turns out \cite{BSQREV} that a
particular one-component equation can represent different variables
in different equations, but in any case only three different one
component equations remain.

The degree growth computations in the present $3\times3$ stencil case
confirm the many results that have been obtained for equations defined
in the $2\times2$ and $2\times3$ stencil and stated in Section
\ref{C:ent}. One additional observation is that it is enough to have
just one non-numeric initial value to differentiate between integrable
and non-integrable equations: for integrable equations the growth is
linear in this case.

\subsection*{Acknowledgements}
I would like to thank C.-M. Viallet for help with the numerical
computations.  I have also benefited from discussions with
B. Grammaticos, G. Gubbiotti, T. Mase, R. Willox, and D.-j. Zhang.

\section*{Appendix}
\subsection*{Integrability by the three-soliton condition}
For the two Hirota bilinear cases we can study integrability by
computing multi-soliton solutions. It is well known that all
one-component Hirota bilinear equations have one- and two-soliton
solutions, but the existence of three-soliton solutions, without
additional constraints, is possible only for integrable equations.

The parameters in equations \eqref{eq:HB7int}
have been chosen so that we have simple one-soliton solutions:
Defining the plane wave factor (corresponding to $e^{kx+\omega t}$) by
\begin{equation}\label{eq:HiPWF}
\rho_{n,m}(k):=c_k\left(\frac{k-1}{k+1}\right)^n\left(\frac{k}{k-1}\right)^m,
\end{equation}
we have the one-soliton solution
\[
\tau_{n,m}=1+\rho_{n,m}(k_1),
\]
and the two-soliton solution
\[
\tau_{n,m}=1+\rho_{n,m}(k_1)+\rho_{n,m}(k_2)+A_{1,2}\,\rho_{n,m}(k_1)\rho_{n,m}(k_2),
\]
where the phase factor is
\begin{equation}\label{eq:Hiphase}
A_{k_i,k_j}:=\frac{(k_i-k_j)^2}{k_i^2+k_ik_j+k_j^2-1}.
\end{equation}
The denominator is typical to Boussinesq type equations.  One can then
verify that \eqref{eq:HB7int} has the three-soliton
solution\footnote{The form of this ansatz is fixed by the requirements
  that if any one of the three solitons goes away the other two
  solitons approach the two-soliton solution above.}
\begin{align*}
  \tau_{n,m}=&1+\rho_{n,m}(k_1)+\rho_{n,m}(k_2)+\rho_{n,m}(k_3)\\
  &+A_{1,2}\,\rho_{n,m}(k_1)\rho_{n,m}(k_2)
+A_{2,3}\,\rho_{n,m}(k_2)\rho_{n,m}(k_3)
+A_{3,1}\,\rho_{n,m}(k_3)\rho_{n,m}(k_1)\\
&+A_{1,2}A_{2,3}A_{3,1}\,\rho_{n,m}(k_1)\rho_{n,m}(k_2)\rho_{n,m}(k_3),
\end{align*}
without any additional restrictions on $k_j$. 

Turning now to the non-integrable case \eqref{eq:HB7non} with the
given coefficients, we have the plane-wave factors
\[
\rho_{n,m}(k):=c_k\left(\frac{-k+3}{9k+3}\right)^n
\left(\frac{-k-1}{k-1}\right)^m,
\]
and then the one-and two-soliton are solutions as above, but with the
phase factor
\[
A_{i,j}=\frac{-3(k_i-k_j)^2}
{[k_ik_j+1][3k_i^2k_j^2-8k_ik_j(k_i+k_j)-3(k_i^2+k_ik_j+k_j^2)]}.
\]
After these an attempt for a three-soliton solution yields the
condition
\[
k_1k_2k_3+k_1+k_2+k_3=0,
\]
and thus the three line-solitons cannot be in general position.
This is typical for the non-integrable case.

\label{lastpage}
\end{document}